# Applying Syntax–Prosody Mapping Hypothesis and Prosodic Well-Formedness Constraints to Neural Sequence-to-Sequence Speech Synthesis

*Kei Furuakwa[1], Takeshi Kishiyama[2], and Satoshi Nakamura[1]*

[1]Nara Institute of Science and Technology, Japan
[2]Graduate School of Arts and Sciences, The University of Tokyo, Japan
`furukawa.kei.fi4@is.naist.jp, kishiyama.t@gmail.com, s-nakamura@is.naist.jp`

## Abstract

End-to-end text-to-speech synthesis (TTS), which generates speech sounds directly from strings of texts or phonemes, has improved the quality of speech synthesis over the conventional TTS. However, most previous studies have been evaluated based on subjective naturalness and have not objectively examined whether they can reproduce pitch patterns of phonological phenomena such as downstep, rhythmic boost, and initial lowering that reflect syntactic structures in Japanese. These phenomena can be linguistically explained by phonological constraints and the syntax–prosody mapping hypothesis (SPMH), which assumes projections from syntactic structures to phonological hierarchy. Although some experiments in psycholinguistics have verified the validity of the SPMH, it is crucial to investigate whether it can be implemented in TTS. To synthesize linguistic phenomena involving syntactic or phonological constraints, we propose a model using phonological symbols based on the SPMH and prosodic well-formedness constraints. Experimental results showed that the proposed method synthesized similar pitch patterns to those reported in linguistics experiments for the phenomena of initial lowering and rhythmic boost. The proposed model efficiently synthesizes phonological phenomena in the test data that were not explicitly included in the training data.

## 1. Introduction

End-to-end TTS systems for alphabetic languages such as English have brought the quality of synthesized speech closer to that of natural speech [1, 2]. In contrast to English, which has 26 different characters, Japanese has a huge number of characters, and the reading of each character is not consistent. Japanese end-to-end speech synthesis attempts to solve this problem by using phoneme sequences [2].

Input to the end-to-end TTS model can be classified into lexical levels, such as those that contribute to lexical distinction, and post lexical levels, which correspond to levels above words such as phonological phrases. In Tokyo Japanese, accents are phonemic because they are elements in addition to the phonemes that contribute to lexical discrimination. The introduction of phoneme sequences and accent symbols as inputs of TTS improves the naturalness of speech synthesis [4, 5]. Fujimoto et al. [6] represented accents as two-dimensional vectors of H and L, which were simultaneously input as one-hot vectors of the phoneme, resulting in improved naturalness of the synthesized speech.

Furthermore, other studies have incorporated information of the post-lexical level, including phonological information, syntactic structure, and syntactic dependency information [6, 7, 8]. Kurihara et al. [6] added prosodic symbols representing not only lexical levels such as phonemes and accent nucleus but also post-lexical prosodic information such as accent phrase boundaries and pauses. As a result, they succeeded in improving the naturalness of the synthesized speech compared to the models whose input is only phonemes. Guo et al. [7] attempted to use syntactic information from parsed trees for end-to-end TTS in English. Syntactic parsing decomposes sentences into syntactic phrase tree structures such as verb phrases and noun phrases. The experimental results of a subjective evaluation show that syntactic features can improve the quality of synthetic speech, especially for longer sentences with complex grammar. Kaiki et al. [8] introduced syntactic symbols of dependency distance to end-to-end TTS for Japanese. As TTS inputs, they adopted prosodic symbols representing the syntactic dependency distances at phrase boundaries; consequently, they observed 1) pause insertion indicating phrase boundaries and 2) F0 resetting at right branch boundaries. These results suggest that implementing prosodic symbols representing syntactic dependency distances enabled syntactic disambiguation of ambiguous sentences.

However, the previous studies have been evaluated based on subjective naturalness and have not objectively examined whether they can reproduce pitch patterns of phonological phenomena in Japanese that are derived from the syntactic origin and phonological constraints.

This study aims to reproduce speech sounds of syntactic and phonological phenomena by applying a syntax–prosody mapping hypothesis and phonological constraints to neural sequence-to-sequence speech synthesis. Although this paper examines phonological phenomena in Japanese, the proposed method is applicable to Japanese and other languages since the theories behind it are proposed to explain the universality of languages.

The rest of this paper is organized as follows. In section 2, we introduce the backgrounds of phonological theories and phonological phenomena. We present the conditions and results of experiments in section 3. In section 4, we provide discussions and concluding remarks.

## 2. Syntax–prosody mapping hypothesis and prosodic well-formedness constraints

In linguistics, most phonological theories of intonation assume phonological structures that are independent of the syntactic structure [9, 10]. It should be assumed that the phonological hierarchy is independent of the syntactic structure because there are phonological phenomena where the same syntactic structure can be created as different prosodies due to phonological constraints. Most theories of Japanese intonation assume two distinct prosodic categories, Minor Phrase (MiP) and Major Phrase (MaP), that can theoretically be unified into recursive PPhrases [11, 12, 13]. MiP is defined by accent culminativity and initial lowering, while MaP is defined as a domain of downstep [11]. In two of these works [10, 11], both the domain of initial lowering and the domain of downstep are considered to be recursive phrases. Ito and Mester [12, 13] defined accent culminativity as an exclusive property of minimal PPhrases, eliminating the need for a theoretical distinction between MaPs and MiPs. By assuming such phonological structures, it is possible to explain the initial lowering and rhythmic boost.

### 2.1. Initial lowering and syntax–prosody mapping hypothesis

The initial lowering is the F0 rise at the beginning of a PPhrase, the realization of a low tone at the left boundary followed by a phrasal high tone in the framework of Autosegmental Metrical (AM) theory [14, 15] and the ToBI transcription system [16]. The boundary tone %L is assigned to the first mora of the PPhrase, and the H tone is assigned to the second, except in cases where the first syllable is accented or the first syllable is heavy [15]. If an initial lowering appears at the beginning of a word, the lowered mora is on the left edge of the PPhrase.

The degree of pitch increase in initial lowering varies in response to syntactic structure [17]. The experimental sentences in the earlier work [17], as shown in Figure 1, consist of sentences that contain the nouns N1, N2, and N3 and have a subject-object-verb (SOV) structure; in tree 1, N1 and N2 constitute the subject noun phrase, and N3 is the object noun phrase; in tree 2, N1 constitutes the subject noun phrase and N2 and N3 constitute the object noun phrase. It was reported that initial lowering is greater at position A before N2 in tree 1, while the initial lowering is greater at position B before N3 in tree 2 [17]. This result can be reinterpreted by the syntax–prosody mapping hypothesis (SPMH) [18]. The SPMH states that syntactic categories (words, phrases, and clauses) be mapped to the corresponding phonological structures (PWords, PPhrases, and PClauses) [18]. The phonological phrase organizations hypothesized from SPMH appear in line b of the figures. Since syntactic edges at position B in tree 1 and at position A in tree 2 are projected onto the edges of multiple PPhrases, the above results can be explained, assuming that the number of edges in the PPhrases is proportional to the degree of the pitch increase in the initial lowering.

### 2.2. Downstep and rhythmic boost

Downstep in Japanese is a pitch range compression triggered by H* + L lexical pitch accents [19, 11]. The F0 peaks of the following words are reached at much lower levels than those of the first accented word. Several studies [11, 19, 20, 21, 22] provide experimental data to indicate that downstep compresses the F0 not only of the second target but also of the third and following words, cumulatively.

Kubozono [21] found that in an accented four-word sequence with a left branch structure, the F0 peak of the third word is reached at the same height or higher than the preceding word. In other words, downstep in the syntagmatic view is canceled at the third word, although this word still exhibits some downstep effect. This phenomenon is called rhythmic boost. A rhythmic boost is observed in the left-branching structure for four-word sequences but not for three-word sequences [23]. Kubozono [21] argued that the left-branching structure as shown on the left side of Figure 2 is prosodically re-phrased as two intermediate PPhrases (MiPs) recursively dominating two minimal PPhrases (PPs) each, as shown on the right side of Figure 2. We believe that the above results could be explained from the viewpoint of initial lowering, assuming that the number of edges in the PPhrases is proportional to the degree of the pitch increase in the initial lowering. This indicates that this phenomenon is independent of syntax and can be explained by the phonological constraints.

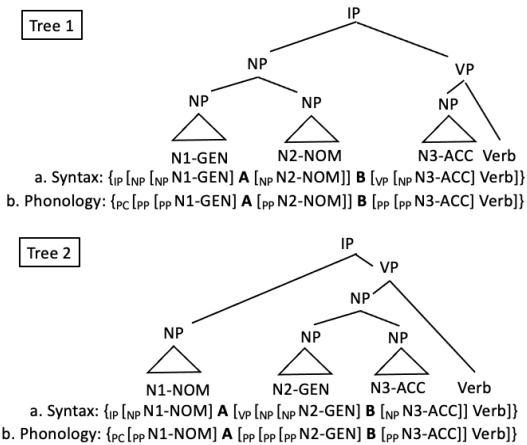

Figure 1: *The syntactic structure of tree 1 and tree 2, and their mapping to the phonological structures. The letters A and B mark the positions of relevant syntactic edges.*

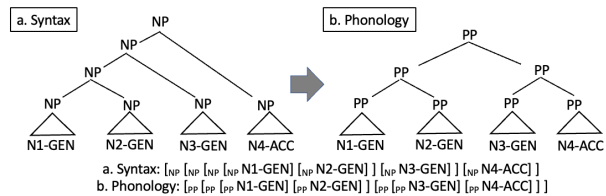

Figure 2: *The syntactic structure of left-branching structure and its mapping to the phonological structure.*

### 2.3. Present study

This study proposes a model in which syntactic structures are mapped to phonological structures via the syntax–prosody mapping hypothesis and prosodic well-formedness constraints. As proposed in phonological theory, syntactic information and phonological information are neutralized in building the phonological structure. In other words, in Guo's and Kaiki's models [9, 10], sentences with the same syntactic structure should have the same syntactic information in the training data, while this cannot be the case in the proposed method.

This study has both engineering and linguistic goals. The engineering goal is to implement the syntax–prosody mapping hypothesis and prosodic well-formedness constraints to reproduce phonological phenomena that are not explicit in the training data. The linguistic goal is to indirectly examine the psychological reality of the knowledge of abstract phonological structures that are not explicit in standard orthography.

## 3. Experiment

We conducted two objective evaluation experiments to investigate the effectiveness of implementing phonological structures in synthesizing phonological phenomena in Tokyo Japanese.

### 3.1. Proposed model and baselines

We propose a model based on phonological structures derived via the SPMH and prosodic well-formedness constraints. In addition to the phoneme symbols, the proposed method uses symbols to indicate accents, PPhrase, and PClause, which are written as "\", "[ ]", and "{ }", respectively. The syntactic structures were automatically obtained from a parser called Haruniwa2 [24]. The outputs of Haruniwa2 were formatted, and verb phrases (VPs) and postpositional phrases (PPs) dominating noun phrase (NPs) were mapped to PPhrases with the symbol "[ ]". For the PClause implementation, we marked IP, the output of Haruniwa2, as PClause with the symbol "{ }"; when there is a CP governing IP, we replaced the CP, rather than IP, with "{ }". Phoneme sequences and accent types were obtained automatically by entering the text into Open Jtalk [25]. According to one work [15], there must be a phonological boundary after every accent owing to accent culminativity and the anti-lapse constraint. In other words, the AA (accented + accented) sequence is separated due to accent culminativity, and the AU (accented + unaccented) sequence is separated due to the anti-lapse constraint. In other words, UA and UU sequences can be a single PPhrase as in [UA] and [UU], while the AA and AU sequences have to be [A][A] and [A][U]. Furthermore, re-phrasing was done so that the right edge of the PPhrase always comes after A[1].

Baseline 1 is a model in which phoneme sequences and accents are the input to TTS. Baseline 1 consists of only the lexical levels that contribute to lexical differentiation: phoneme sequences and accents. The difference between Baseline 1 and the proposed method is a phonological hierarchy that is not explicitly indicated in the orthography.

Baseline 2 is the proposed model 1 in Kaiki et al. [10]. The model uses symbols indicating accents, initial lowering, and syntactic dependency distance to the phoneme symbols. The symbols indicating accents and initial lowering are written as "\" and "^", respectively. As for the dependency distance, "#1" to "#6" are used. For example, in tree 1, *wagashiyasan-no* (the Japanese confectioner's) modifies the bean-seller one position behind, so its dependency distance is #1. On the other hand, in tree 2, *wagashiyasan-ga* (the Japanese confectioner) is the subject, and the dependent verb is three words behind, so its dependency distance is expressed as #3. If the dependency length is more than 6, the distance is expressed as #6. In Baseline 2, boundary markers are determined solely from syntactic information, whereas they are determined from the interaction of phonological and syntactic constraints in the proposed method.

### 3.2. Speech database and procedure

We used the speech database created by Kaiki et al. [10] in this experiment. The database consists of an oral transcription of the Arabian Nights and its reading voice by a single speaker: 190 texts, with average durations of 8 minutes and 21 seconds, total 26 hours and 26 minutes [26]. The speech data were automatically divided into sentence units by CTC Segmentation and phoneme alignment, resulting in a dataset of 11,615 sentences excluding error sentences [10]. The sentences were automatically converted to the model of the proposed method using Haruniwa2 and Open Jtalk. A total of 5,953 sentences were used as the dataset, excluding the sentences with errors. From the input sequences, Japanese Tacotron 2 [27, 28] generated a mel-spectrum, which is converted to waveforms via Griffin-Lim in ESPNet2 [29]. Of the 5,953 sentences, 5,453 sentences were used for training, and 250 each were used for validation and testing.

---

[1] The text data for replication will be available when the paper is accepted.

### 3.3. Objective evaluation 1: degree of initial lowering reflecting syntactic structures

We conducted an objective evaluation experiment on initial lowering reflecting syntactic structures.

#### 3.3.1. Experimental items and measurements

This objective evaluation experiment targets the same sentence structures as that shown in Figure 1 [17]. An example of an item from the experiment is shown in Table 1. The underlined moras *waga*, *mame*, and *memo* are the locations where initial lowering should occur in N1, N2, and N3, respectively.

Table 1: *Example sentences and test data for objective evaluation 1.*

| tree 1 | N1 | A | N2 | B | N3 | |
|---|---|---|---|---|---|---|
| item | waga shiyasan-no | | mame uriyaku-ga | | memo gaki-o | moraimashita. |
| tone | LHHHHHH | | LHHHHHH | | LHHHH | LHHHH*LL |
| gloss | Japanese confectioner-GEN | | been.seller-NOM | | notes-ACC | received |
| | 'The Japanese confectioner's bean-seller received some notes.' | | | | | |
| tree 2 | N1 | A | N2 | B | N3 | |
| item | waga shiyasan-ga | | mame uriyaku-no | | memo gaki-o | moraimashita. |
| tone | LHHHHHH | | LHHHHHH | | LHHHH | LHHHH*LL |
| gloss | Japanese confectioner-NOM | | been.seller-GEN | | notes-ACC | received |
| | 'The Japanese confectioner received the bean-seller's notes.' | | | | | |

**baseline 1 (phonems and accents)**
 tree 1: w a g a sh i y a s a N n o m a m e u r i y a k u g a m e m o g a k i o m o r a i m a \ sh
 tree 2: w a g a sh i y a s a N g a m a m e u r i y a k u n o m e m o g a k i o m o r a i m a \ sh
**baseline 2 (phonemes, accents, initial lowering, and dependency length)**
 tree 1: w a ^ g a sh i y a s a N n o #1 m a ^ m e u r i y a k u g a #2 m e ^ m o g a k i o #1 m o ^ r a i m a \ sh I t a .
 tree 2: w a ^ g a sh i y a s a N g a #3 m a ^ m e u r i y a k u n o #1 m e ^ m o g a k i o #1 m o ^ r a i m a \ sh I t a .
**proposed (phonemes, accents, and phonological structures)**
 tree 1: { [ [ w a g a sh i y a s a N n o ] [ m a m e u r i y a k u g a ] ] [ [ m e m o g a k i o ] [ m o r a i m a \ sh I t a ] ] . }
 tree 2: { [ w a g a sh i y a s a N g a ] [ [ [ m a m e u r i y a k u n o ] [ m e m o g a k i o ] ] [ m o r a i m a \ sh I t a ] ] . }

In Baseline 1, the difference between tree 1 and tree 2 is limited to the phonemes. In Baseline 2, #1 is inserted at position A and #2 at position B in tree 1, while #3 is inserted at position A and #1 at position B in tree 2, reflecting syntactic dependency distances. Since the proposed model implements the SPMH, the number of PPhrases is the same as in Figure 1. The number of "[" indicating the left edges of the PPhrases changes according to the difference of the trees.

Sound files were annotated using Praat [29]. Segmentation between the moras was done on the basis of formants and waveforms. The following measurements were taken: (i) RiseSizeA = the maximum pitch of the second mora after A minus the minimal pitch of the first mora after A in semitones;

(ii) RiseSizeB = the maximum pitch of the second mora after B minus the minimal pitch of the first mora after B in semitones.

### 3.3.2. Results

The results are shown in Table 2. In natural prosody, RiseSizeA is greater than RiseSizeB in tree 1, while RiseSizeB is greater than RiseSizeA in tree 2 [17]. The rows in Table 2 that do not follow the same pattern as the natural speech are grayed out. In the proposed model and Baseline 2, initial lowering is greater at position A than at position B in tree 1, while the initial lowering is greater at position B than at position A in tree 2. As a whole, the proposed model and Baseline 2 showed the same pattern as the natural prosody reported earlier [17].

Table 1: *F0 range of initial lowering at A and B.*

| model | sentence | cond | RiseSizeA | RiseSizeB | Same pattern as natural prosody? |
|---|---|---|---|---|---|
| baseline 1 | 1 | tree 1 | 0.68 | -0.26 | No |
| baseline 2 | 1 | tree 1 | 1.75 | 12.12 | Yes |
| proposed | 1 | tree 1 | 8.17 | 11.84 | Yes |
| baseline 1 | 1 | tree 2 | 0.72 | 0.51 | Yes |
| baseline 2 | 1 | tree 2 | 14.17 | 3.50 | Yes |
| proposed | 1 | tree 2 | 11.96 | 1.56 | Yes |
| baseline 1 | 2 | tree 1 | 2.17 | 8.06 | Yes |
| baseline 2 | 2 | tree 1 | 4.32 | 10.58 | Yes |
| proposed | 2 | tree 1 | 6.12 | 9.39 | Yes |
| baseline 1 | 2 | tree 2 | 0.50 | -2.27 | No |
| baseline 2 | 2 | tree 2 | 12.35 | 2.71 | Yes |
| proposed | 2 | tree 2 | 9.30 | 9.07 | Yes |

### 3.4. Objective evaluation 2: rhythmic boost

We conducted an objective evaluation experiment on rhythmic boost.

### 3.4.1. Experimental items and measurements

This objective evaluation experiment targets the sentence structures used in Shinya et al. [23], as shown in Figure 2. An example of an item from the experiment is shown in Table 3.

Table 3: *Example sentences and test data for objective evaluation 2.*

| input | N1 | N2 | N3 | N4 | | | |
|---|---|---|---|---|---|---|---|
| item | kinou | yamanashi-no | moriguchi-no | aniyome-no | waruguchi-o | kouen-de | tsutaeta |
| tone | LHH | LH*LLL | LH*LLL | LH*LLL | LH*LLL | HHHHH | LHHHH |
| gloss | yesterday | Yamanashi-GEN | Moriguchi-GEN | sister.in.law-GEN | bad things-ACC | park-in | tell |

'Yesterday, I said the bad things about the sister-in-law of Moriguchi in Yamanashi in the park.'

baseline 1 (phonems and accents)
  4N: k i n o o y a m a n a s h i n o m o r i g u c h i n o a n i y o m e n o
       w a r u g u c h i o k o o e N d e t s U t a e t a .
baseline 2 (phonemes, accents, initial lowering, and dependency length)
  4N: k i / n o o # 6 y a / m a \ n a s h i n o # 1 m o / r i \ g u c h i n o # 1 a / n i \ y o m e n o
       # 1 w a / r u \ g u c h i o # 2 k o / o e N d e # 1 t s U / t a e t a .
proposed (phonemes, accents, and phonological structures)
  4N: { [ k i n o o ] [ [ [ [ y a m a \ n a s h i n o ] [ m o r i \ g u c h i n o ] ] [ a n i \ y o m e n o ]
       [ w a r u \ g u c h i o ] ] [ k o o e N d e ] [ t s U t a e t a ] ] . }

In natural speech, pitch range is compressed by downstep from N1 to N2 and from N3 to N4, but the effect of downstep is weakened by the rhythmic boost from N2 to N3 [23]. In Baseline 2, #1 is inserted from N1 to N4 since the N1 to N4 have a left-branching structure. In the proposed model, recursive PPhrases (e.g., additional "] [" between N2 and N3) were inserted to coalesce N1 and N2, N3 and N4 to implement the re-phrasing of phonological structures proposed by Kubozono [21] as in Figure 2. Note that this re-phrasing for the texts does not exist in the training data.

The basic measurement procedures were identical to those of the evaluation 1. The following measurements were taken: (i) N1-N2 = maximum F0 of N1 minus maximum F0 of N2 in semitones; (ii) N2-N3 = maximum F0 of N2 minus maximum F0 of N3 in semitones; (iii) N3-N4 = maximum F0 of N3 minus maximum F0 of N4 in semitones.

### 3.4.2. Results

The results are shown in Table 4. The pitch contours of sentence 1 from Baseline 2 and from the proposed model are shown in Figure 3. In natural speech, the measurements of N1-N2 and N3-N4 are negative because of downstep, while N2-N3 approaches zero or becomes positive because of rhythmic boost. The rows in Table 4 that do not follow the same pattern as the natural speech are grayed out. In the proposed model, only the same patterns as those of natural language were observed, while this is not the case in Baselines 1 and 2.

Table 4: *F0 descent between each noun.*

| model | sentence | N1-N2 | N2-N3 | N3-N4 | Same pattern as natural prosody? |
|---|---|---|---|---|---|
| baseline 1 | 1 | 4.95 | -4.02 | 1.57 | No |
| baseline 2 | 1 | 2.25 | 0.53 | 1.00 | No |
| proposed | 1 | -3.79 | -0.41 | -3.98 | Yes |
| baseline 1 | 2 | 1.29 | -2.71 | -2.24 | No |
| baseline 2 | 2 | 0.62 | 2.60 | -4.54 | No |
| proposed | 2 | -2.84 | 1.94 | -1.12 | Yes |

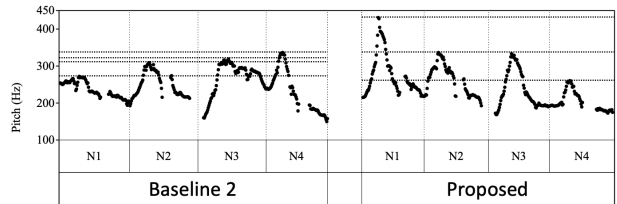

Figure 3: *Pitch contours of sentence 1 from Baseline 2 (left) and from the proposed model (right).*

## 4. Discussion and Conclusions

The proposed model and Baseline 2 synthesized speech with patterns similar to natural speech for initial lowering, reflecting syntactic structures reported in [17]. In addition, while the output patterns of Baseline 1 and Baseline 2 are not constant in terms of rhythmic boost, the proposed model consistently reproduces the natural patterns of rhythmic boost. By merging and neutralizing syntactic and phonological constraints as phonological structures proposed in linguistics, we believe that the proposed method was able to reproduce not only syntactic but also phonological phenomena.

The contribution of the proposed method from the viewpoint of engineering is that it efficiently synthesizes phonological phenomena in the test data that were not explicitly included in the training data. Sentences with four accented words with uniformly left-branching structure are rare in the database, and no re-phrasing of recursive PPhrases as proposed in Kubozono [21] was performed in the training data. As a linguistic contribution, the fact that the model reproduces untrained phonological phenomena by explicitly providing phonological structure indirectly suggests the validity of the psychological reality of the phonological structures. However, since the principles of neural net TTS are different from those of human speech, this proposal needs further research toward improving the psychological reality of computational models.

## 5. Acknowledgements

The author is grateful to Nobuyoshi Kaiki and Yuki Yano for the assistance and helpful discussions. Part of this work is supported by JSPS KAKENHI Grant Number JP21H05054.

# 6. References

[1] N. Li, S. Liu, Y. Liu, S. Zhao, M. Liu, and M. Zhou, "Close to human quality TTS with transformer," arXiv:1809.08895, 2018.

[2] J. Shen, R. Pang, R. J. Weiss, M. Schuster, N. Jaitly, Z. Yang, Z. Chen, Y. Zhang, Y. Wang, R. Skerrv-Ryan, R. A. Saurous, Y. Agiomvrgiannakis, and Y. Wu, "Natural TTS synthesis by conditioning wavenet on Mel Spectrogram predictions," *2018 IEEE International Conference on Acoustics, Speech and Signal Processing (ICASSP)*, 2018.

[3] Y. Yasuda, X. Wang, S. Takaki, and J. Yamagishi, "Investigation of enhanced Tacotron text-to-speech synthesis systems with self-attention for pitch accent language," in *Proceedings of ICASSP*, 2019, pp. 6905–6909.

[4] T. Fujimoto, K. Hashimoto, K. Oura, Y. Nankaku, K. Tokuda "Impacts of Input Linguistic Feature Representation on Japanese End-to-End Speech Synthesis," Proc. of 10th ISCA Speech Synthesis Workshop (SSW), pp.166-171, Vienna, Austria, Sep. 2019

[5] K. Kurihara, N. Seiyama, and T. Kumano, "Prosodic features control by symbols as input of sequence-to-sequence acoustic modeling for neural tts," *IEICE Transactions on Information and Systems*, vol. E104.D, no. 2, pp. 302–311, 2021.

[6] K. Kurihara, N. Seiyama, T. Kumano, and A. Imai, "Study of Japanese End-to-End speech synthesis method that inputting kana and prosodic symbols," in *Autumn Meeting of the Acoustical Society of Japan*, pp. 1083–1084, 2018, (in Japanese).

[7] H. Guo, F. K. Soong, L. He, and L. Xie, "Exploiting syntactic features in a parsed tree to improve end-to-end TTS," *Interspeech 2019*, 2019.

[8] N. Kaiki, S. Sakti, and S. Nakamura, "Using local phrase dependency structure information in neural sequence-to-sequence speech synthesis," *2021 24th Conference of the Oriental COCOSDA International Committee for the Co-ordination and Standardisation of Speech Databases and Assessment Techniques (O-COCOSDA)*, 2021.

[9] J. Pierrehumbert, "The phonology and phonetics of English intonation," Cambridge, MA: MIT dissertation, 1980.

[10] E. Selkirk, "Phonology and syntax: The relation between sound and structure," Cambridge, MA: MIT Press, 1984.

[11] S. Ishihara, "14 syntax–phonology interface," *Handbook of Japanese Phonetics and Phonology*, pp. 569–618, 2015.

[12] J. Ito, and A. Mester, "Recursive prosodic phrasing in Japanese," *Prosody matters: Essays in honor of Elisabeth Selkirk*, pp. 280-303, 2012.

[13] J. Ito, and A. Mester, "Prosodic subcategories in Japanese," *Lingua*, *124*, pp. 20-40, 2013.

[14] J. Pierrehumbert, and M. Beckman, "Japanese tone structure," Cambridge: MIT Press, 1988.

[15] Y. Igarashi, "13 Intonation," *Handbook of Japanese Phonetics and Phonology*, pp. 525–568, 2015.

[16] J. J. Venditti, "The J_TOBI model of Japanese intonation," *Prosodic Typology*, pp. 172–200, 2005.

[17] E. Selkirk, T. Shinya, and M. Sugahara, "Degree of initial lowering in Japanese as a reflex of prosodic structure organization," Proc. 15th ICPhS, Barcelona, Spain, 491-494, 2003.

[18] E. Selkirk, "The syntax-phonology interface," *The Handbook of Phonological Theory*, pp. 435–484, 2011.

[19] W. J. Poser, "The phonetics and phonology of tone and intonation in Japanese," *Journal of Japanese Linguistics*, vol. 12, no. 1, pp. 208–209, 1990.

[20] H. Kubozono, "The Organization of Japanese Prosody," Ph.D. dissertation, Edinburgh University, 1988.

[21] H. Kubozono, "Syntactic and rhythmic effects on downstep in Japanese," *Phonology*, *6*(1), pp. 39-67, 1989.

[22] S. Ishihara, "Japanese downstep revisited," *Natural Language & Linguistic Theory*, vol. 34, no. 4, pp. 1389–1443, 2016.

[23] T. Shinya, E. Selkirk, and S. Kawahara, "Rhythmic boost and recursive minor phrase in Japanese," In *Speech Prosody 2004, International Conference*, 2004.

[24] S. W. Horn, A. Butler, and K. Yoshimoto, "Keyaki Treebank segmentation and part-of speech labelling," In Proceedings of the 23th Meeting of the Association for Natural Language Processing, pages 414–417, 2017.

[25] *Open JTalk*. [Online]. Available: http://open-jtalk.sourceforge.net/. [Accessed: 20-Mar-2022].

[26] Takehazuchi, "On the reading of relieving stories Arabian Nights oral translation," https://o-keil.com/okinu-ba-ba/wordpress/?p=818 (in Japanese)

[27] Y. Wang, R. J. Skerry-Ryan, D. Stanton, Y. Wu, R. J. Weiss, N. Jaitly, Z. Yang, Y. Xiao, Z. Chen, S. Bengio, Q. Le, Y. Agiomyrgiannakis, R. Clark, and R. A. Saurous, "Tacotron: Towards end-to-end speech synthesis," *Interspeech 2017*, 2017.

[28] S. Watanabe, T. Hori, S. Karita, T. Hayashi, J. Nishitoba, Y. Unno, N. Enrique Yalta Soplin, J. Heymann, M. Wiesner, N. Chen, A. Renduchintala, and T. Ochiai, "ESPnet: End-to-end speech processing toolkit," *Interspeech 2018*, 2018.

[29] P. Boersma, and D. Weenink, "Praat: A System for Doing Phonetics by Computer" Version 6.0. 39, 2018.